\documentclass[aps,a4paper,twocolumn,superscriptaddress,showpacs,nofootinbib,floatfix]{revtex4}

\usepackage{amsmath}
\usepackage{amsfonts}
\usepackage{amssymb}
\usepackage{bm}
\usepackage{latexsym}
\usepackage[dvips]{graphicx,color}
\usepackage{float}

\newcommand{\Omegat}{\mbox{$\underline{\underline{\mathbf{\Omega}}}_t$}}
\newcommand{\Omegato}{\mbox{$\underline{\underline{\mathbf{\Omega}}}_{t_0}$}}
\newcommand{\OmegaTto}{\mbox{$^{T}\underline{\underline{\mathbf{\Omega}}}_{t_0}$}}
\newcommand{\OmegaTtdelta}{\mbox{$^{T}\underline{\underline{\mathbf{\Omega}}}_{\Delta^{\ast}}$}}
\newcommand{\Omegatildet}{\mbox{$\underline{\underline{\mathbf{\widetilde{\Omega}}}}_t$}}
\newcommand{\Omegatildeto}{\mbox{$\underline{\underline{\mathbf{\widetilde{\Omega}}}}_{t_0}$}}
\newcommand{\OmegatildeTto}{\mbox{$^{T}\underline{\underline{\mathbf{\widetilde{\Omega}}}}_{t_0}$}}
\newcommand{\rsfig}[1]{
  \begin{center}
    \vspace{0.3cm}
    \includegraphics*[width=8.5cm]{{#1}}
  \end{center}
}
\newcommand{\rrsfig}[2]{
  \begin{center}
    \vspace{0.3cm}
    \includegraphics*[width=0.46\textwidth]{{#1}}
    \hfill{}
    \vspace{0.3cm}
    \includegraphics*[width=0.46\textwidth]{{#2}}
  \end{center}
}

\newcommand{\comment}[1]%
{\textsf{\textcolor{red}{#1}}}

\begin{document}
\sloppy

\title{Evolution of entanglements during the response to a uniaxial
  deformation of lamellar triblock copolymers and polymer glasses}

\author{F.~L\'eonforte}
\email[Email: ]{leonforte@theorie.physik.uni-goettingen.de}
\affiliation{Institut f\"ur Theoretische Physik, Georg-August-Universit\"at,
  Friedrich-Hund-Platz 1, 37077 G\"ottingen, Germany}

\begin{abstract}
  Using coarse-grained molecular-dynamics simulations, a generic
  styrene-\emph{block}-butadiene-\emph{block}-styrene ($SBS$) triblock
  copolymer under lamellar conformation is
  used in order to investigate the mutual entanglement evolution when a 
  structure of alternating glassy ($S$)/rubbery ($B$) layers is submitted to an imposed deformation. 
  By varying the
  amount of \emph{loop} chains between each phase, i.e. \emph{noncrossing}
  chains, it is possible to
  generate different types of $S/B$ interface definitions. A specific boundary driven tensile
  strain protocol has been developed in order to mimic
  "real" experiments and measure the stress-strain curve. 
  The same protocol is also applied to a reference state consisting
  in a directed glassy homopolymers, as well as to an isotropic glassy polymer.
  The evolution of initial mutual entanglements from the undeformed samples during the whole
  deformation process is monitored. It
  is shown for all considered systems that initial
  entanglements mostly participate to the preyield regime of the
  stress-strain curve and that this network is debonded during the strain-hardening regime. 
  For triblocks with a
  non-null amount of \emph{crossing} chains, the
  lower the amount is, the longer the memory effect of the
  initial entanglement network in the postyield regime is. \emph{On 
  the fly} distributions of entanglements, which depart from the
  postyield regime, depict memory effects and long time correlations 
  during the strain-hardening regime. For triblocks,
  \emph{loop} chains reinforce these effects.
\end{abstract}

\pacs{
  61.41.+e Polymers, elastomers, plastics;
  82.20.Wt Computational modeling, simulations;
  82.35 Jk Copolymers, phase transitions, structure;
  82.35.Lr Physical properties of polymers.
}

\maketitle

\section{Introduction}
\label{sec:intro}

The macroscopic properties of multiphase polymers involve more than the simple
addition of the individual properties of each of their constituents. They
are the result of alchemy between the material morphology (phase
repartitions and regularity of this repartition), the properties 
of each constituent, and the characteristics of the interfaces or the
manner continuity between phases is assured. In 
styrene-\emph{block}-butadiene-\emph{block}-styrene ($SBS$) triblocks copolymers
of glassy/rubbery/glassy type under lamellar morphology \cite{Seguela81,Cohen00},
chains are indeed constituted by two phases linked by covalent bonds, with each phase having
its own mechanical properties. The way the phases are
``connected'' at the microscopic level, i.e., at their interfaces, may thus have an impact on the
energetically involved transfer mechanisms at larger scales. For instance, a chain effectively
forming a bridge between two $S$ domains will transfer the stress more
efficiently than a chain forming a \emph{loop} between $S$ and $B$, except if the behavior of the
latter is modified by the presence of trapped entanglements.

Coarse-grained Molecular Dynamics (MD) simulations are well suited to bridge the scales 
between global mechanical properties of the material and those of each phases, including
coarse-grained scale details for the nature of their interfaces. At
global scale, the main features of the mechanical behavior of lamellar
$SBS$ materials under elongation can be summarized in a linear elastic response with
significant rigidity at low deformation up to a yield point, a
yielding and drawing at higher deformations with possible alteration
of lamellar morphology into a chevronlike one \cite{Seguela81}, 
and the possibility to reach
very large elongations. Many other features can also be noticed
concerning cyclic strain deformations and unloading
\cite{Cohen00}. Large deformations have a strong impact on the
morphology of thermoplastic elastomers, affecting the
nanostructures. In $SBS$ triblocks, several studies have pointed out
the presence of spherical polystyrene domains in the matrix \cite{Kaelble73,Chen77}, playing
the role of fillers \cite{Holden69}, as well as the role of entanglements in this
large strain regime. In MD simulations, such domain effects are
hard to reproduce because one has to simulate huge systems with several
repeated lamellar units, while entanglement effects are well
accessible.

In glassy/rubbery/glassy triblocks, the behavior of entanglements is
quite complex to relate to. In the rubbery phase, these ones are
transient and may be mapped onto
slip-links that slide along the chains during deformation, then
allowing chains to explore many conformations. In contrary, in the
glassy domains, entanglements are trapped and chains slow down and may visit
different conformations under strong enough applied deformation. 
This leads to stress dissipation and activated disentanglements.
Additionally, in realistic triblocks, one has to take into account the
contribution of \emph{loop} chains, as well as diblock chains, for which
the dynamics and stress transmission under deformation may be altered
by the presence of frozen entanglements. 

In this paper, we analyze the evolution of entanglements in $SBS$ triblocks and polymer glasses, when these materials are submitted to a macroscopic deformation. Unlike theoretical models \cite{Rubinstein02,Rubinstein97,Uchida08} that take as input time- and deformation-independent entanglement sets, we analyze how entanglements evolve and depart from this assumption. Paper is then organized as follows. In Sec.~\ref{sec:technical}, technical details about simulation model and methods are first given. The primitive path analysis \cite{Everaers04, Sukumaran05} (PPA) is used to identify the entanglements during the whole deformation process. Details about the method are briefly given and how this one is extended to give a statistical representation of the entanglements. The mechanical response to a uniaxial deformation of the simulated systems is discussed in Sec.~\ref{sec:Response}. Then, the Sec.~\ref{sec:StatPPA} deals with mutual entanglements statistics and how their statistical representation can be related to the response behavior at larger scales. Initial entanglements set from the undeformed state are then monitored during the whole deformation process, as well as the ``dynamical'' or \emph{on the fly} contribution of entanglements activated by the deformation. Finally, in Sec.~\ref{sec:Conclude}, concluding remarks are given, and outlooks are discussed.

\section{Polymer model and methods}
\label{sec:technical}

In the following, MD simulations are performed using a well-established coarse-grained model \cite{Kremer}. 
The polymer is treated as a chain of $N = \sum_{\alpha} N_{\alpha}$ beads (where $\alpha$ denotes the species 
for block copolymers), which we refer to as monomers, of mass $m=1$ connected by a spring to
form a linear chain. The beads
interact with a classical $n$-species Lennard-Jones excluded volume interaction,

\begin{equation}\label{LJpot}
  \mathrm{U^{\alpha\beta}_{LJ}(r)} =
  4u^0_{\alpha\beta}\left[\left(a_{\alpha\beta}/r\right)^{12} - \left(a_{\alpha\beta}/r\right)^6\right] 
\end{equation}

\noindent For particle distances, $r > r_c=2.5a$, the potential is cut-off and shifted such that it is continuous at $r_c$.
Indices $(\alpha,\beta)$ stand for the different types of pairs, $S-S$, $B-B$, and $S-B$ for $SBS$ triblocks, and $u^0_{\alpha\beta}$ is in units of $u^0$. Adjacent monomers along the chains are coupled 
through the well-known anharmonic finite extensible nonlinear elastic potential (FENE) in addition to Eq.\eqref{LJpot},

\begin{equation}\label{FENEpot}
  \mathrm{U_{FENE}(r)} = -0.5 k R^2_0 \ln{\left(1 - \left(r/R_0\right)^2\right)}
\end{equation}

\noindent where model parameters are identical to those given in Ref.~\cite{Kremer}, namely 
$k=30u^0/a^2$ and $R_0=1.5a$, chosen so that unphysical bond 
crossings and chain breaking are eliminated. All quantities will be
expressed in terms of the molecular diameter $a\equiv 1$, binding energy
$u^0\equiv 1$ and characteristic time $\tau_{LJ}=\sqrt{ma^2/u^0}$. Under such conditions, the glass
transition temperature for the $\alpha=\beta=1.0$ model is sited around $k_B T_g\sim 0.42u^0$.

Newton's equations of motion are integrated with the velocity-Verlet
method and using a time step $\delta t=0.006\tau_{LJ}$. Simulation cells of size
$(L_x,L_y,L_z)$ are filled with $M$ chains of size $N$.

\subsection{Simulated systems}
\label{subsec:triblock}

\begin{figure}
  \begin{center}
    \includegraphics*[width=8.5cm]{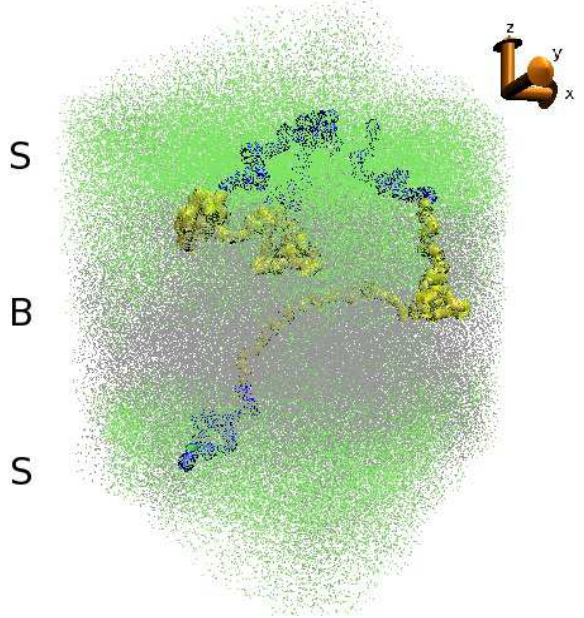}
    \caption{\label{SBS.fig}
      (Color online) Snapshot of a $PS/PB/PS$ lamellar triblock before deformation is applied. 
      The glassy phases are labeled with $S$, while rubbery region is labeled
      with $B$. Evanescent points are used in background to represent monomers from each part.
      A \emph{loop} chain (left) and a \emph{crossing} chain (right) from the same sample are also superimposed.}
  \end{center}
\end{figure}

The considered triblocks copolymers are intended to mimic lamellar $SBS$
triblocks such as the one depicted in Fig.~\ref{SBS.fig}, 
in the sense of alternating glassy/rubbery layers. Thus, our simple model 
consists in polymer chains composed of $2N_S=N_B=100$ monomers, with a total 
length per chain $N=200$. The total temperature being fixed to
$k_BT=0.3u^0_{\alpha\beta}$, the glassy phase is simulated using a value
of $u^0_{SS}=1.0u^0$, and the rubbery phase with $u^0_{BB}=0.5u^0$. 
In that way, the effective temperature in the rubbery phase is $k_B T_B\sim 0.6u^0 > k_B T_g$.
Finally, the incompatibility between each phase is controlled by setting $u^0_{SB}=10^{-2}u^0$, in order
to maintain the lamellar morphology in the strong segregation regime.

Triblocks are prepared using an enhanced version of the radical-like polymerization method
\cite{Perez}, which is an extension of Gao's work \cite{Gao95}. We then
refer to Refs.~\cite{Perez, Gao95} for further details concerning this method.

This one has then been applied in order to generate four
types of triblocks according to the percentage of \emph{loop} chains, namely
$0\%$, $30\%$, $60\%$ and $100\%$. These \emph{loop} chains are defined as chains that 
do not cross the entire material from a glassy phase $S$ to the opposite $S$ and by the way that
also do not cross the whole rubbery phase $B$. Then, they loop back to their starting glassy phase 
while a part of the chain lies in the rubbery region. Such a chain is depicted in Fig.~\ref{SBS.fig}.

For each system, the number of
obtained chains of desired length $N=200$ is respectively $M=192$,
$M=188$, $M=194$ and $M=191$, and systems are kept under a 
$\langle P\rangle =0$ state after a run of $10^6$ MD steps. 
Finally, monomers dynamic in each phases
has been checked by computing the incoherent intermediate scattering
function $F_s(\mathbf{k},t)=\langle
\frac{1}{N_{\alpha}}\sum_{j=1}^{N_{\alpha}}
\exp{\left(i\mathbf{k}.\delta\mathbf{r}_j(t-t_0)\right)}\rangle$ where
$\delta\mathbf{r}(t-t_0)=\mathbf{r}(t)-\mathbf{r}(t_0)$, which depicts
(not shown) a typical glassy behavior \cite{Bennemann99} in $S$ domains, and a
rapid relaxation behavior in the rubbery $B$ domain.

In order to have a reference state, a homopolymer is also simulated
using the $0\%$ configuration and setting
$u^0_{\alpha\beta}=u^0_{SS}\equiv 1.0u^0$ this for all $(\alpha,\beta)$. A
subsequent simulation run is performed during $10^7$ MD steps at
$k_BT=0.3u^0_{SS}$ and $\langle P\rangle=0$ using the same
anisotropic Nos\'e-Hoover barostat. After this run, the
density of chain ends was checked to correspond to that of a directed homopolymer,
namely, that chain ends are located in $z\sim 0$ and $z\sim L_z$. In
the following, this directed homopolymer will be referred to as $hP$.

Finally, an isotropic homopolymer has also been generated by annealing
the previous $hP$ polymer at $k_BT=1.0u^0_{SS}$ and $\langle
P\rangle=0.5$ during $2\times10^7$ MD steps, and cooling it to
$k_BT=0.3u^0_{SS}$ for a time of $10^6$ MD steps. Then
an additional run of $10^6$ MDS at the same temperature and
pressure $\langle P\rangle=0$ has been performed. In this case, 
the density of chain ends was
checked to be spatially homogeneous. In the following, we will
refer to this sample as $iP$ polymer.

Glassy homopolymers and glassy phases in triblocks 
are then out of equilibrium. The $\alpha$ relaxation time $\tau_{\alpha}$ is defined as the time
at which $F_s(\mathbf{k}_1,\tau_{\alpha})=1/e$, where $\mathbf{k}_1$ is the first peak position
in the structure factor. We found $\tau_{\alpha}$ of same order
than the typical time used to perform the deformation experiment. In that sense, ageing effects may not significantly contribute to the glassy dynamics upon deformation, and are expected to not alter the mechanical response of the 
materials.

\subsection{Boundary driven uniaxial tensile test}
\label{subsec:uniaxial}

A method in which the
deformation is applied first at boundaries and then is self-transmitted to
the sample has been developed in order to mimic experiments \cite{Makke}. 
The periodic boundary condition in the $z$-axis (direction of the applied
deformation) is removed, while the other directions are kept
periodic. Hence, chains that cross the periodic-image 
boundary in the $z$ direction were unwrapped and cut. 
This induces length polydispersity and
then slightly decreases the total number of monomers per samples.
For instance, it only alters the glassy parts of the triblocks.
Another simulation run of $5\times10^6$ MD steps is added to ensure all
samples are still under an overall $\langle P\rangle=0$ state.

Grips are defined in the two glassy regions with a thickness
of $2.6a_{SS}$, the upper one for $z > L_z - 2.6a_{SS}$ and the lower
for $z < 2.6a_{SS}$. Then, forces and velocities are set to zero and
$v_0$ respectively for all monomers lying in the grips, requesting a
$v_0=0$ for lower grip. The lower grip is then coupled to an elastic
spring with a force
$\mathbf{F}^l=-K\left(\mathbf{r}^l_{cm}(t)-\mathbf{r}^l_{cm}(0)\right)$
adjusted so that the average lower grip velocity is null, as requested. The upper
grip is also coupled to an elastic spring with a force also adjusted
to preserve the requested upper grip velocity $v_o\mathbf{e_z}$, and
is given by $\mathbf{F}^u=-K\left(\mathbf{r}^u_{cm}(t) -
  \mathbf{r}^u_{cm}(0) - v_0t\mathbf{e_z}\right)$. In both
expressions, $\mathbf{r}_{cm}(t)$ denotes the center of mass position
of grips at time $t$.

During this boundary driven deformation protocol, a Berendsen thermostat is used with a heat
bath of $k_BT=0.3u^0_{\alpha\beta}$, while an anisotropic
Nos\'e-Hoover barostat is applied in order to keep a lateral pressure
$P_{xx}=P_{yy}=0$. The obtained deformation thus mimics a uniaxial
tensile test. As discussed in Ref.~\cite{Makke}, the value of $K$
should be chosen as ten times as stiff as the initial sample, in order
to lead to a numerically stable deformation scheme; such a condition is
fulfilled in our numerical experiment.

\begin{table}[h!]
  \begin{tabular}{|c||c|c|c|c||c|c|}
    \hline
    sample code & $E_Y$  & $G_r$   & $\sigma_{flow}$ & $\langle N_e\rangle$\\
    \hline\hline
    $0\%$       & $2.77$  & $0.033$ & $0.21$          & $47$\\
	\hline
	$30\%$      & $2.81$  & $0.044$ & $0.22$          & $33$\\
	\hline
	$60\%$      & $2.59$  & $0.042$ & $0.23$          & $26$\\
	\hline
	$100\%$     & $2.12$  & $0.005$ & $0.17$          & $9$\\
	\hline
	$hP$        & $13.4$ & $0.27$  & $0.42$          & $60$\\
	\hline
	$iP$        & $11.8$ & $0.15$  & $0.40$          & $27$\\
    \hline
  \end{tabular}
  \caption{\label{infos.tab} Young's modulus $E_Y$, hardening modulus
    $G_r$, and post-yield flow stress $\sigma_{flow}$ from
    stress-strain curve in Fig.~\ref{stress_strain.fig}, for all
    simulated systems. Also given, the average entanglement length $\langle
    N_e\rangle$ for undeformed samples and from the primitive path analysis.}
\end{table}

\subsection{Primitive Path Analysis}
\label{subsec:PPA}

Several theories try to explain the deformation of polymer melts in both
rubbery and glassy states. In the rubbery state, polymers can explore many conformations under the
constraint of cross-link points (entanglements, covalent bonds...). Entropic-based
classical theory of rubber elasticity \cite{Arruda1, Arruda2, Boyce}, which neglects entanglements, 
may be used. However, in the glassy state, chains are slow down and may visit
conformations in a noncontinuous way under deformation. This leads to stress dissipation effects 
\cite{Bocquet09, Hoy07, Hoy08} under deformation, noisy term currently not present in
rubber elasticity theories.

Statistical mechanic theories improve the lack of cross-link
effects by including entanglements to the
constitutive relation via some additional terms \cite{Everaers96,Rubinstein02}, as well as
by developing subnetwork deformation formalisms \cite{Rubinstein97, Uchida08}.
It's then supposed that entanglements depart from an initial distribution, while nothing more
is said about its evolution upon deformation. This is a crucial point, as nothing compels polymer 
glasses and thermoplastic triblocks to not explore other sets of entanglement distributions.

\begin{figure}[h!]
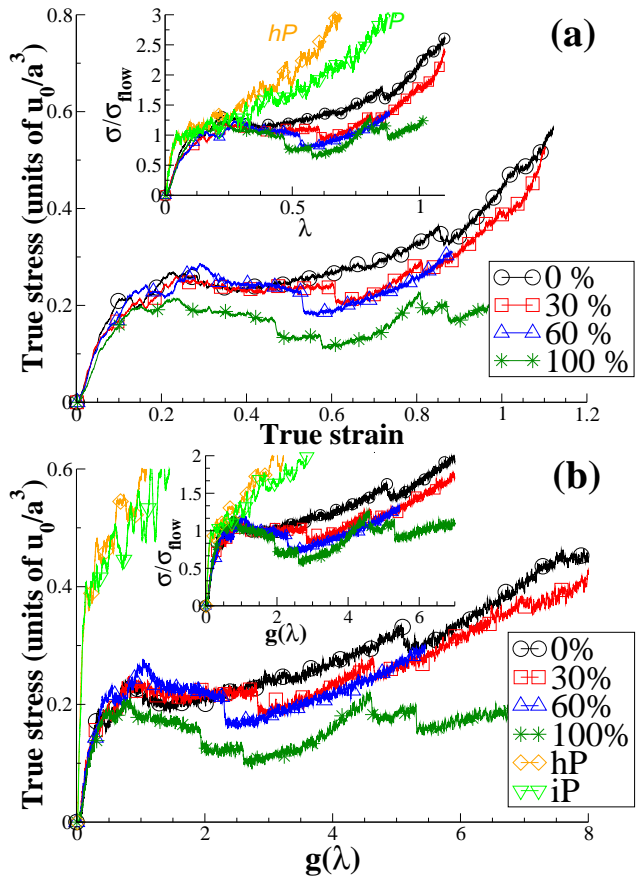

  \rrsfig{figure2a.eps}{figure2b.eps} \caption{(Color online) True
    stress $\sigma_{zz}$ measured during boundary driven tensile test
    against {\bf (a)} true strain $\lambda=L_z(t)/L_z(0)$, and {\bf (b)} Gaussian
    strain $g(\lambda)=\lambda^2-\lambda^{-1}$. \emph{Inset:}
    Gaussian strain-hardening plot, in which stresses are normalized by $\sigma_{flow}$.}
  \label{stress_strain.fig}
\end{figure}

Under deformation, the stress-strain relation of these materials is expected to follow a
superposition principle, with one contribution from inherent glassy
structures and their contribution to energy landscape, leading to the
so-called yield stress, and the other contribution from
strain-hardening effects. Events from the latter may be related to
thermally activated processes \cite{Haward93,Govaert04,Sarva07,Boyce90}, 
conformational effects \cite{McKechnie93}, or topological
entanglement effects \cite{Vorselaars09,Fetters94,Fetters99,Lyulin05}. It was also 
shown in \cite{Hoy06,Uchida08} for a wide range of polymer glasses,
that underlying entanglement effects contribute to the strain hardening.

The PPA framework \cite{Everaers04, Sukumaran05} 
furnishes a practical tool in order to identify the entanglements.
Such an analysis has been performed on undeformed
triblocks and $hP$ and $iP$ polymer glasses, averaging the
number of monomers $N_e$ per Kuhn segment over all chains. Results are
summarized in Tab.~\ref{infos.tab}, in which it first appears for
triblocks that $\langle N_e\rangle$ decreases with the increasing amount of \emph{loop}
chains, thus leading on the same manner to an increase of the
entanglement density $\rho_e=\rho/2\langle N_e\rangle$. For polymer glasses, the entanglement density
$\rho_e$ is higher for the isotropic $iP$ glass than for the directed
$hP$ glass. This is also expected regarding to the average imposed
chains conformations for the latter.

In order to study more precisely the evolution of entanglements during the tensile test numerical experiment, 
the PPA has been used every $10^4$ MD steps on dumped configurations, and $10^4$ times during the whole 
deformation process. For each dumped configuration to which PPA has been applied (hereinafter referred to PP chains), 
three quantities are extracted:

\begin{itemize}
\item[$\lbrack 1\rbrack$] for each PP polymer chain, the curvature between a monomer $i$ and its 
	neighbors $j=i-\delta/2$ and $k=i+\delta/2$ is computed. The spatial position and monomer $Id$ 
	where the bending angle is maximal is then stored. We refer to this ensemble as $MaxBend$, and
	$\delta=4$ is used in the following.
\item[$\lbrack 2\rbrack$] according to a distance criterion $r_l\equiv\mu.r_c$, each monomer of a chain $1$ having another
  	monomer of a chain $2$ at a distance less than $r_l$ is an element
  	of the ensemble $Contacts$ for chains $1$ and $2$.
\item[$\lbrack 3\rbrack$] a more restrictive ensemble per chain is built as the intersection 
	of both previous ensembles. Entanglements taking part of this ensemble can be considered as $Mutual$ entanglements.
\end{itemize}

Both $Contacts$ and $Mutual$ ensembles depend on the value
of the contact length $r_l=\mu.r_c$, with
$\mu\in\rbrack 0,\infty\lbrack$ and $r_c$ the cut-off distance in
Eq.\eqref{LJpot}. Several values were tested. For $\mu < r_c^{-1}$, the distance criterion to build contacts
is too restrictive, and the ensemble contains too few elements to perform a correct
statistical analysis. In the limit of $\mu\rightarrow\infty$, one has
$\mathcal{P}_{a^{+}/a^{-}}(\mu,t,t_0)\rightarrow 0,\,\forall(t,t_0)$ (see Sec.~\ref{subsubsec:apm}),
as expected in this mean-field limit.
We found optimal results for $\mu=r_c^{-1}$, and checked that varying it do not alter main results. This value
will be used in the following.

Finally, from a computational point of view, the complete analysis, including parallel implementation of PPA on a $16$ processors computer and well optimized post-processing codes, took around $1300\,s$ CPU per dumped configuration.

\section{Response to deformation}
\label{sec:Response}

A uniaxial tensile test is
performed at a temperature $k_BT=0.3u^0_{\alpha\beta}$ under
anisotropic barostatic conditions $P_{xx}=P_{yy}=0$, with an applied
upper grip velocity $v_0\mathbf{e_z}=10^{-4}a_{\alpha\beta}/\tau_{LJ}\mathbf{e_z}$. 
The true stress $\sigma_{zz}$ is
monitored as a function of the elongation
$\lambda=L_z(t)/L_z(0)$. Results are shown in
Fig.~\ref{stress_strain.fig}(a) for all samples and versus the true
strain $ln{(\lambda)}$. One can separate the stress-strain curve in
three regions: \emph{(i)} an elastic regime with possible
localized irreversible monomer rearrangements \cite{Tanguy06},
\emph{(ii)} a maximum $\sigma_Y$ which marks the onset of plasticity, directly
followed by a strain softening regime, \emph{(iii)} a quasi-ideal
plastic flow $\sigma_{flow}$ and a strain
hardening regime at large deformations.
For triblocks, the first stage extends to strain less than $5-7\%$,
while for $hP$ and $iP$ polymer glasses, this one covers the region
with strain less than $2\%$. From this elastic regime, the Young's
modulus $E_Y$ is extracted, with higher values for polymer
glasses compared to ones for triblocks. Results are given in Tab.~\ref{infos.tab}.

The ratio of Young's moduli
for homopolymers to triblocks is in quantitative agreement with one
observed experimentally \cite{Cohen00}. We also observe that Young's
modulus for directed polymer glass $hP$ is naturally higher than the
one for the isotropic glass $iP$, for a deformation imposed along the $hP$
preferential chains average orientation. Depending the percentage of
\emph{loop} chains, different stiffness are observed for triblocks in the elastic regime.
It appears that the $100\%$ \emph{loop} seems to be weaker, while it is hard to conclude
between other triblocks. Larger systems with several lamellae are needed to go further.

In Fig.~\ref{stress_strain.fig}(b), the true stress is also
plotted in the Gaussian strain-hardening framework using the relation
$\sigma=\sigma_{flow}+G_rg(\lambda)$, where
$g(\lambda)=\lambda^2-\lambda^{-1}$. Fitting the stress-strain curves
with this expression defines the hardening modulus $G_r$ and the
offset yield stress $\sigma_{flow}$. Results are
summarized on Tab.~\ref{infos.tab} for all systems. This kind of plot
allows us to separate the plastic flow and stress-hardening regions
for $g(\lambda)\gtrsim 1$,
to the preyield one for $g(\lambda) \lesssim 0.8$. By this way, it
also appears for triblocks that stress-strain curves seem to show two peaks around the
yield zone.

In Figs.~\ref{stress_strain.fig}(a) and (b), we note different behaviors under
deformation for triblocks and pure polymer glasses. The directed $hP$
glass displays a stronger strain-hardening than the isotropic $iP$
glass. Then polymer glasses have a stronger strain-hardening than
triblocks. For triblocks, large stress drops appear suddenly and earlier when 
the amount of \emph{loop} chains increases.

The presence of the elastomer also affects the response of triblocks. 
Instead of breaking, drops are
nearly followed by local strain hardening zones, until another large drop
appears, thus relaxing the stress in the whole material. This has an
impact on chains orientation, mainly in the glassy phase, where a
chevron-type structural change may occur \cite{Huy03}. 

\begin{figure}
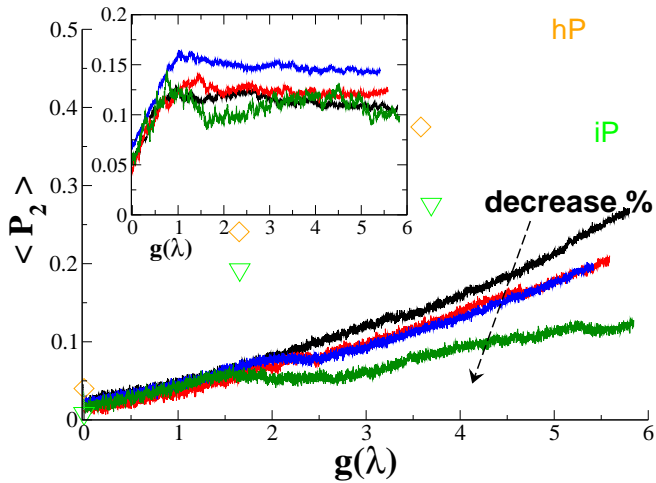

  \rsfig{figure3.eps} \caption{(Color online) Mean second Legendre
    polynomial $P_2$ between bond segments and applied deformation
    director $\textbf{e}_z$ upon deformation. $\langle P_2\rangle$ is plotted
    for triblocks, for chains in the rubbery phase, and on (inset),
    for chains in the glassy phase. Both plots are against the Gaussian strain
    $g(\lambda)=\lambda^2-\lambda^{-1}$. Also plotted is the Legendre
    polynomial for $hP$ and $iP$ homopolymers. The increase of
    $\langle P_2\rangle$ mimics the trend of polymer chain segments to align
    along to $\textbf{e}_z$, the direction of applied deformation.}
  \label{P2.fig}
\end{figure} 

The orientation process of chains upon deformation can be monitored by the second Legendre polynomial
$P_2=\frac{1}{2}\left(3\cos^2(\theta) - 1\right)$, where $\theta$ is
formed by the vector between monomers $\lbrack i,i+1\rbrack$ of the same chain
and the direction of the applied deformation. The average orientation of the
segments is taken as the ensemble average $\langle P_2\rangle$, where
$\langle P_2\rangle=1$ means a tendency of chains to orient along the
direction of the applied deformation, and $\langle P_2\rangle=0$ means
an average random orientation of chains.

In Fig.~\ref{P2.fig}, $\langle P_2\rangle$
is plotted, for triblocks, for monomers in the rubbery phase. 
Average chain orientation becomes weaker with the
increase of the amount of \emph{loop} chains. This is expected because of 
the average isotropic conformations of these chains. In the inset of the 
Fig.~\ref{P2.fig},
an orientation process is activated in the glassy phase, until the yield point
is reached around $g(\lambda)\sim 1$. After yield, glassy chains adopt an
average different orientation that does not change anymore 
and with a strength that increases with the amount of \emph{loop} chains.
This tilting process is associated to a topological
transition of conformations of glassy chains, due to the
presence of discontinuities in the elastic medium of triblocks at the interfaces. 
The energy transferred to the material under the applied deformation leads to a
localization of stresses (and/or specific monomer displacements) in the
vicinity of the interfaces. This energy transferred to the glassy
phase is strong enough after yield stress to cause an instability of the
lamellae, leading to a transition from a lamellar to a chevron
type structure \cite{Read99}.

\begin{figure}[h!]
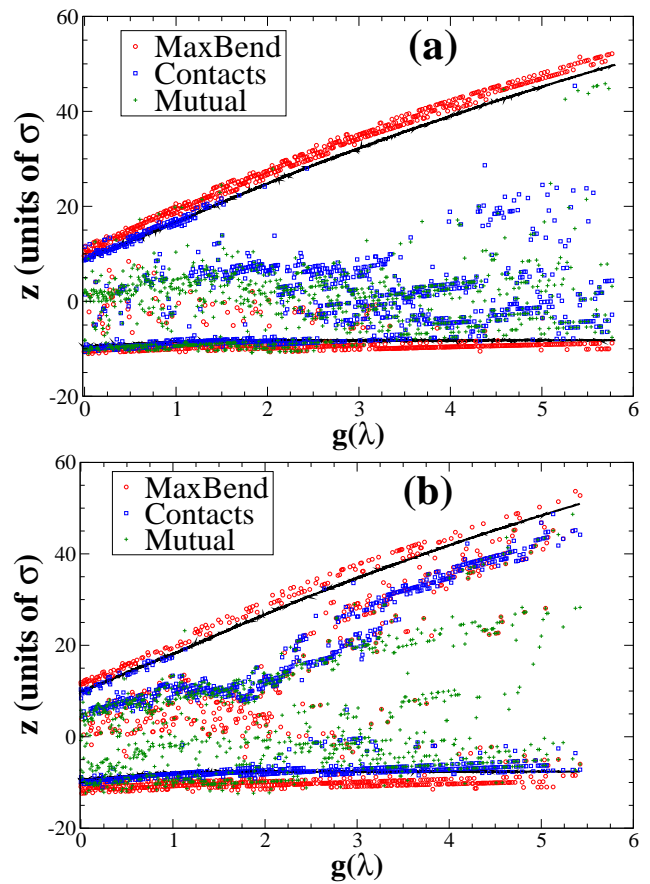

  \rrsfig{figure4a.eps}{figure4b.eps} \caption{(Color online)
    Spatially resolved distributions of entanglements from each
    ensemble (see Sec.~\ref{subsec:PPA} for details) during deformation. Black lines
    represent the average position of the interfaces $S/B$ (for $z < 0$) and
    $B/S$ (for $z > 0$), and symbols are related to
    the considered ensemble. {\bf (a)} For $0\%$ \emph{loop}
    and {\bf (b)} for $60\%$ \emph{loop} triblocks.}
  \label{distEntgl.fig}
\end{figure}

We note that the $30\%$ and $60\%$ \emph{loop} triblocks have a tendency to orient more
in the glassy phase than the $0\%$ triblock. Thus the tilting process is more pronounced when
the amount of \emph{loop} chains increases, meaning that more energy is transferred trough the interface to
the glassy phase. For the special case of the $100\%$ \emph{loop}, large fluctuations are observed due to
the higher capacity of chains to reorient. Furthermore, it clearly appears that the changes in trend of 
$\langle P_2\rangle$ in the rubbery phase can be well mapped onto the large stress drops in the stress-strain
curve of Fig.~\ref{stress_strain.fig}(b). Hence, these ones find their origin in the release of topological
constraints and average chain orientation.

In Figs.~\ref{distEntgl.fig}(a) and (b), the time and space analysis of
entanglements from ensembles $MaxBend$, $Contacts$ and $Mutual$,
are plotted during the whole deformation process. In these figures, the black lines
represent the average position of the interfaces between $S/B$ phases for $z
< 0.0$, and $B/S$ phases for $z>0.0$. For the $0\%$ \emph{loop}
triblock in Fig.~\ref{distEntgl.fig}(a), $MaxBend$ entanglements are
mostly localized in the glassy regions, and in the vicinity of the interfaces. For
the $60\%$ \emph{loop} triblock in Fig.~\ref{distEntgl.fig}(b), another contribution appears
in the rubbery phase, until the stress drop located at
$g(\lambda)\sim 2$ (see Fig.~\ref{stress_strain.fig}) is reached and
after which $MaxBend$ entanglements localize on both parts of the
interfaces. $Contact$ entanglements probe the dynamics of relevant contacts between chains upon
deformation. For both triblocks, relevant contacts disappear in the vicinity of the upper 
moving interface, nearby to the value of the yield point at $g(\lambda)\sim 1$. 
This lost of the interchain connectivity affects the balance with maximum
curvature entanglements, then leading to a local interface morphology
change, i.e. the chevron-type orientation transition. 

For the $0\%$ \emph{loop} triblock, these contacts persist in the
rubbery phase and far from the moving interface. On the other way, they tend to be 
spatially distributed at a fixed distance from the moving interface
for the $60\%$ \emph{loop} triblock, and coincide with some maximum bending entanglements.
This might be an effect of the \emph{loop} chains, which
preserve the material stiffness and stress transmission by adapting their conformations. It explains
why strain-hardening for $30\%$ and $60\%$ \emph{loop} triblocks is weakly dependent of the amount
of \emph{loop} chains, in regard to the $0\%$ \emph{loop} triblock. On the other way, it is found (not shown)
for $hP$ and $iP$ homopolymers that the same entanglement sets are well distributed over the whole simulation box. 

Finally, we note that the ensemble $Mutual$, which is more restrictive, seems to be a good
compromise in order to capture relevant entanglements. In
the following, we will mainly focus on that ensemble.

\section{Entanglements statistics}
\label{sec:StatPPA}

Using statistical ensembles for entanglements defined in the previous
Sec.~\ref{subsec:PPA}, a specific attention is paid on the
``participative dynamics'' of these entanglements upon the tensile test numerical
experiment depicted in Sec.~\ref{sec:Response}. To this aim, we split
the contribution of these entanglements in two main contributions:
from the initial distribution, and from the distributions \emph{on the fly}.

\subsection{Contribution of initial entanglements}
\label{subsec:InitStatPPA}

The initial distributions of entanglements form the initial networks of triblocks and polymer
glasses. For their representation, we define the
matrix $\Omegat$ of distributions of events from the $Mutual$ ensemble set at
time $t$. This matrix has $M$ rows and each row represents a chain of $N$ monomers.
Element of a row is equal to $0$ when the monomer $i$ is not an element of the
considered ensemble and $1$ otherwise.

\subsubsection{Creation \emph{vs} annihilation}
\label{subsubsec:apm}

We first define $a^{+}(\mu,t,t_0)$ as the number of events present at time $t$ but
not at time $t_0$ divided by the size of the ensemble set at time $t$
times the amount of time $(t-t_0)$, thus
characterizing the rate of created events at this time. Using the
above representation: 

\begin{equation}\label{aplus_rate}
  a^{+}(\mu,t,t_0) = \frac{1}{(t-t_0)N_t}\mathrm{Tr}\left(\Omegat\cdot\OmegatildeTto\right)
\end{equation}

\noindent where $N_t$ is the size of the ensemble set at time $t$ and
$\OmegatildeTto$ contains complementary elements (Boolean $NOT$) with respect to the
universe $(0,1)$ for each state defining a monomer in the transposed
matrix $\OmegaTto$ of $\Omegato$. Hence, the operation 
$\widetilde{\mathcal{O}}:\Omegato\longrightarrow\Omegatildeto$ is related to
the Boolean $NOT$ transformation applied to elements of each row. 
In practice, if a monomer of a chain has a state $0$ in a row
of $\Omegato$ at $t_0$, then such a monomer will have a value $1$ in
the column of the associated transposed matrix $\OmegatildeTto$.

\begin{figure}
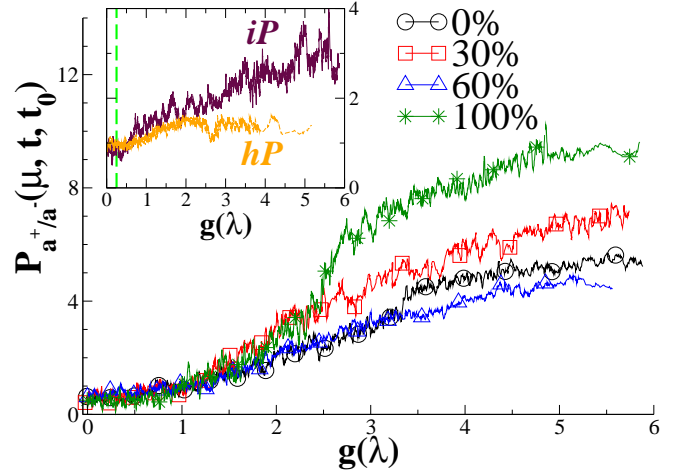

  \rsfig{figure5.eps} \caption{(Color online) Relative ratio $\mathcal{P}_{a^{+}/a^{-}}(\mu,t,t_0)$ of
    created \emph{vs} annihilated entanglements during the whole deformation
    process. Events are taken from the initial $Mutual$ entanglements set. 
    \emph{Main panel:} results for triblocks. \emph{Inset:}
    same ratio for the $iP$ and $hP$ polymer glasses. The dashed line
    represents the yield point for these glasses.}
  \label{rate_ap-am_init.fig}
\end{figure} 

Conversely, we define
$a^{-}(\mu,t,t_0)$ as the number of events present at time $t_0$ but
not at time $t$ and also divided by the size of the ensemble set at
time $t$ times the time length $(t-t_0)$, thus characterizing the
rate of annihilated events at measuring time. In a more compact
notation, one has:

\begin{equation}\label{aminus_rate}
  a^{-}(\mu,t,t_0) = \frac{1}{(t-t_0)N_t}\mathrm{Tr}\left(\Omegatildet\cdot\OmegaTto\right)
\end{equation}

\noindent where $\Omegatildet$ contains complementary elements (Boolean $NOT$) with
respect to $(0,1)$ for each state defining a monomer in the matrix
$\Omegat$. Thus, if a monomer of a chain has a state $0$ in a row of $\Omegat$, it will become
$1$ in the same row of $\Omegatildet$.

The quantities Eq.\eqref{aplus_rate} and Eq.\eqref{aminus_rate}
may differ during the deformation, leading to an asymmetric
creation-annihilation process. The ratio
$\mathcal{P}_{a^{+}/a^{-}}(\mu,t,t_0)$ captures
such a process.

In Fig.~\ref{rate_ap-am_init.fig}, the variation
of the ratio $\mathcal{P}_{a^{+}/a^{-}}(\mu,t,t_0)$ is plotted for the
$Mutual$ ensemble set of entanglements. For triblocks, and before yield at $g(\lambda)\sim 1$, 
balance between alive and created events is achieved
with regard to the initial entanglements set. Then after yield,
more and more new events are created,
which unbalances the ratio with annihilated ones, then leading to an
asymmetric creation-annihilation process of entanglements.
It coincides with the chevron-type transition in the inset of Fig.~\ref{P2.fig} that allows
glassy chain ends to release some topological constraints.
The asymmetric process has an higher magnitude in the case of the $100\%$ \emph{loop} sample, 
with changes in slope that also correspond to the large drops appearing in the stress-strain 
curve in Fig.~\ref{stress_strain.fig}. 

After yield, these more created or activated events can be seen as
the result of ``cascades'' of disentanglements of older events, where such processes alter chain
conformation and, this, in order to maintain the rigidity of the
system. It means that the $100\%$ \emph{loop}
triblock is more affected by the deformation in the plastic flow
regime than other triblocks and needs to explore more chain conformations in order to 
preserve a mechanical stability.

Intriguingly, it appears that the $30\%$ \emph{loop} triblock follows the same trend when 
compared to the $60\%$ \emph{loop} one. The latter is indeed closer to the $0\%$ \emph{loop} 
behaviour. A lower balance is then achieved for this triblock, which means that it is not forced 
to explore more and more new sets and destroy older ones in order to preserve its mechanical 
stability. In that sense, this amount of \emph{loop} chains seems to much preserve the triblock integrity, 
and plays in favour of the future metastable states during the deformation.
This can be related to the more pronounced chevron-type process for this triblock, which compensates the
loss of topological constraints due to the low amount of \emph{crossing} chains. A this point, a more
precise study is needed to explore this compensation effect, and how this one depends on the 
amount of \emph{loop} chains.

In the inset of Fig.~\ref{rate_ap-am_init.fig}, the ratio
$\mathcal{P}_{a^{+}/a^{-}}(\mu,t,t_0)$ is plotted for the
$iP$ and $hP$ polymer glasses. Before yield, balance between created and
annihilated events is observed, followed after yield by an asymmetric
process. Its magnitude for the $iP$ glass is higher than the one for the directed $hP$ glass, because 
new metastable topological intermediate configurations are more accessible for $iP$. Despite of the glassy state,
and under deformation, energetic barriers have then an higher probability
to be crossed, leading to this higher ratio.
Finally, the ratio $\mathcal{P}_{a^{+}/a^{-}}(\mu,t,t_0)$ for $iP$ and $hP$ glasses
is also weaker than the one for triblocks. It is mainly due to the presence of a rubbery 
region in triblocks, in which chains can explore more conformations.

\subsubsection{Survival events}
\label{subsubsec:ps}

Additionally, we define the distribution $\mathcal{P}_s(\mu,t,t_0)$ of $Mutual$ survival events between
initial time $t_0$ and the measuring time $t$ during deformation. 
Using the same formalism for
Eqs.\eqref{aplus_rate} and \eqref{aminus_rate}, it can be written:

\begin{equation}\label{survival_init}
  \mathcal{P}_s(\mu,t,t_0) = \frac{1}{N_t}\mathrm{Tr}\left(\Omegat\cdot\OmegaTto\right)
\end{equation}

\begin{figure}
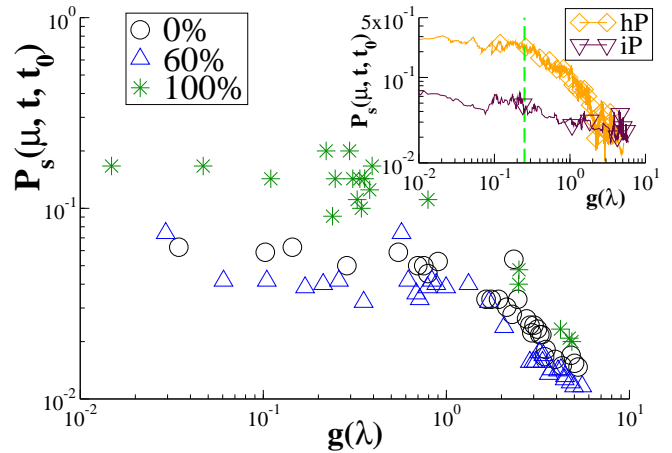

  \rsfig{figure6.eps} \caption{(Color online) Distribution
    $\mathcal{P}_s(\mu,t,t_0)$ of survival
    events between initial undeformed state $t_0$ and deformation time $t$.
    Events are part of the $Mutual$ ensemble
    set of entanglements, and data are plotted for
    triblocks. The $30\%$ \emph{loop} chains triblock is omitted for clarity.
    \emph{Inset:} same quantity for $hP$ and $iP$ polymer
    glasses. The vertical dashed line marks the onset of plastic flow.}
  \label{surv_init.fig}
\end{figure}

Such a distribution is plotted in Fig.~\ref{surv_init.fig}. It clearly appears two distinct 
regimes with increasing $g(\lambda)$: a plateau until yield strain is reached,
followed by a power law decrease of survival events
$\mathcal{P}_s(\mu,t,t_0)\sim g(\lambda)^{-\beta}$. For triblocks,
$\beta\approx 0.83,\,0.74,\,0.68,\,0.97$ for $0,\,30,\,60$ and $100\%$
\emph{loop} chains respectively. Hence, for the three first
triblocks, the plateau value before yield is quantitatively the same.
Except the $100\%$ \emph{loop} triblock, the rate of loss
of the contribution of initial mutual entanglements after yield decreases
with the increase of the amount of \emph{loop} chains. The triblock with
the lowest amount explores more rapidly new
entanglements set on deformation than ones with a higher amount; 
\emph{loop} chains offer a longer memory effect of the initial mutual entanglements network.

The plateau can be interpreted as the fact that
initial $Mutual$ entanglements mainly contribute to
the elastic linear (and nonlinear) part of the strain-stress curve,
while their contribution to the plastic flow and strain-hardening regime
decreases with increasing strain. An higher preyield ``plateau'' is observed
for the $100\%$ \emph{loop} chain triblock, in which chains more preserve the 
initial network of mutual entanglements, but on counterpart 
make the material stiffer and more
sensible to high deformations. An higher postyield power-law exponent is also
measured for this triblock, which means that this
material needs to explore new states more rapidly.

Finally, in the inset of Fig.~\ref{surv_init.fig}, Eq.~\eqref{survival_init} is
plotted for the $hP$ and $iP$ glasses. Due to their chain average orientation, the preyield
regime depicts a higher survival magnitude for the directed glass than for the isotropic one.
In the postyield regime, the power law behavior is fitted with an exponent
$\beta\approx 1.01$ for $hP$ polymer glass, and $\beta\approx 0.2$
for $iP$ glass. The rate of loss of initial mutual entanglements is then higher for the directed glass
than for the isotropic one. It is because, given the initial $hP$ glass entanglement network, the
chain average orientations are too constraining to expect a mechanical equilibrium in the postyield regime.
This glass needs to explore new mutual entanglement states more rapidly upon deformation than the isotropic one.

\subsection{Contribution of \emph{on the fly} entanglements}
\label{subsec:DeltaStatPPA}

The participative dynamics of mutual entanglements is
evaluated by monitoring the contribution of their distributions
to the current deformation time $t$, given a spanning time $\Delta^{\ast}$. 
To this aim, we define the retarded distribution $\mathcal{P}_s(\mu,\Delta^{\ast})$ of
survival events between time $t$ and a new origin $\widetilde{t_0}=t -
\Delta^{\ast}$. This distribution can be written using the same formalism as for
Eq.\eqref{survival_init}:

\begin{equation}\label{survival_delta}
  \mathcal{P}_s(\mu,\Delta^{\ast}) = \frac{1}{N_t}\mathrm{Tr}\left(\Omegat\cdot\OmegaTtdelta\right)
\end{equation}

\begin{figure}
  \begin{center}
    \includegraphics*[width=8.6cm]{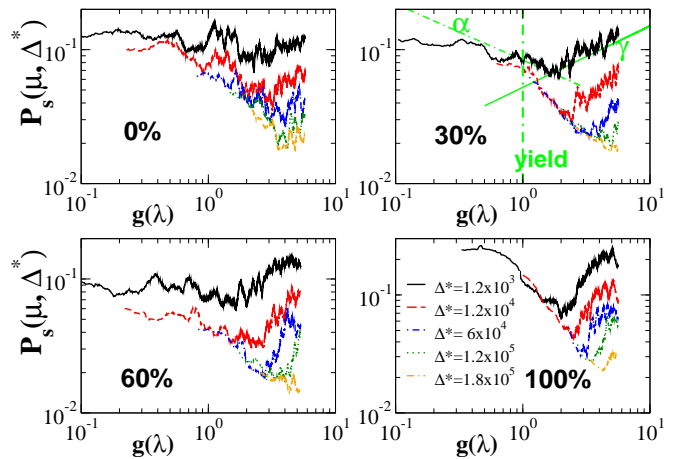}
    \caption{(Color online) Distribution
      $\mathcal{P}_s(\mu,\Delta^{\ast})$ of survival
      events between deformation times $t$ and $t - \Delta^{\ast}$.
      Events are part of the $Mutual$ ensemble
      set of entanglements, and distributions are plotted for
      triblocks with specified percentage of \emph{loop}
      chains. The parameter $\Delta^{\ast}$ is given in $\tau_{LJ}$ units.
      Exponents $\alpha$ and $\gamma$ that characterize the two regimes
      discussed in the text are also drawn. The vertical dashed line marks the 
      yield point.
    } \label{surv_delta_SBS.fig}
  \end{center}
\end{figure}

In Eq.\eqref{survival_delta}, the quantity $\OmegaTtdelta$ is then
related to the matrix of distribution of mutual entanglements at
a given origin $\Delta^{\ast}$. The distribution $\mathcal{P}_s(\mu,\Delta^{\ast})$ is plotted in
Fig.~\ref{surv_delta_SBS.fig} for the different $SBS$ triblocks and
their amount of \emph{loop} chains, i.e. from $0\%$ to $100\%$, versus
the Gaussian strain $g(\lambda)=\lambda^2 - \lambda^{-1}$ and
$\lambda=L_z(t)/L_z(0)$. 

In Fig.~\ref{surv_delta_SBS.fig}, the ``probe'' times
$\Delta^{\ast}$ are given in $\tau_{LJ}$ units, with all
$\Delta^{\ast} < t_{tot}/2$ and the total amount
of simulation time $t_{tot}=6\times10^5\tau_{LJ}$ for all simulated
systems. Values of the probe times are taken according to specific
deformation zones of the stress-strain curves in Fig.~\ref{stress_strain.fig}. 
For the elastic zone, where $t_{elastic}$ is an estimate
of the upper time for the elastic limit, we have $\Delta^{\ast}=1.2\times10^3,\, 1.2\times10^4 <
t_{elastic}$. The value $\Delta^{\ast}=6\times10^4$ is chosen to lie between $t_{p1}$ and $t_{p2}$,
which are estimates of the deformation
time associated with the two peaks in the yielding zone of the
stress-strain curve. Finally, both values $\Delta^{\ast}=1.2\times
10^5,\,1.8\times10^5 > t_{hard}$, where $t_{hard}$ is the time after which
strain-hardening occurs.

It appears in Fig.~\ref{surv_delta_SBS.fig} before yield and $\Delta^{\ast}<t_{elastic}$,
that distributions have a lower magnitude when the amount of
\emph{loop} chains increases, except for $100\%$ \emph{loop}. If \emph{loop} chains maintain the
initial set of entanglements, the $0\%$ triblock is more stable than other triblocks
with regard to new \emph{on the fly} entanglement sets. The $100\%$ \emph{loop} triblock
depicts higher distributions, which is a direct effect of the special
amount of these chains as already discussed in Sec.~\ref{subsec:InitStatPPA}.

The $\mathcal{P}_s(\mu,\Delta^{\ast})$ curves seem to follow a master curve
which can be split in three parts: \emph{(1)} a plateau
for $\Delta^{\ast}<t_{elastic}$, which stays until the yield
strain at $g(\lambda)\sim 1$ is reached, \emph{(2)} a power law decrease fitted
with an exponent $\alpha$, and \emph{(3)} a power low
increase also expressed in terms of an exponent $\gamma$. The transition
from the $\alpha$ to $\gamma$ regime gives rise to a minimum, which we
denote $\mathit{min}\lbrace \mathcal{P}_s(\mu,\Delta^{\ast})\rbrace$. 

Events in the $\alpha$ regime are
related with initial events from the preyield zone, and quantify their contributions to
events moving away from the yield point. The minimum
$\mathit{min}\lbrace \mathcal{P}_s(\mu,\Delta^{\ast})\rbrace$ is
related to the contribution of events at yield point to postyield
ones. Finally, the $\gamma$ regime characterizes the contribution of
events from the post-yield region to ones in the strain-hardening regime.
Hence, an increase of the contributions in this regime marks the emergence 
of long-time correlations and memory effects.

\begin{figure}
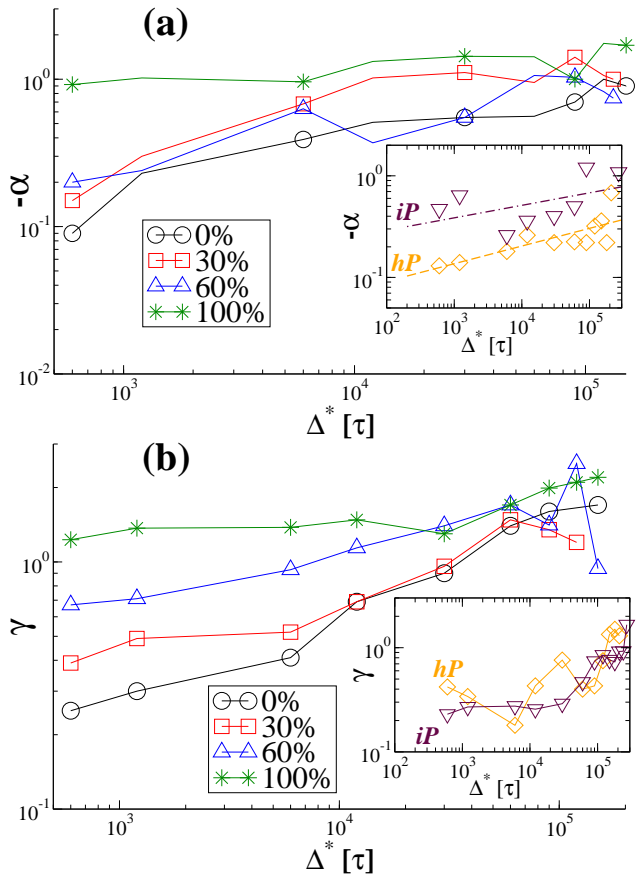

  \rrsfig{figure8a.eps}{figure8b.eps} \caption{(Color online)
    Variation of $\alpha$ \textbf{(a)} and $\gamma$ \textbf{(b)} 
    exponents depicted in Fig.~\ref{surv_delta_SBS.fig} against the probe time 
    $\Delta^{\ast}$
   	that separates two measuring times upon deformation. 
   	Exponents characterize the two power-law behaviors for
    $\mathcal{P}_s(\mu,\Delta^{\ast})$. On both main panels,
    data are plotted for triblocks, and on \emph{insets} for $hP$
    and $iP$ polymer glasses.}
  \label{slopes_surv.fig}
\end{figure}

In the $\gamma$ regime, the distributions of survival mutual entanglements separated by the 
shortest $\Delta^{\ast}$ have a higher magnitude than ones for longer
$\Delta^{\ast}$. On counterpart, the route to the memory effect is
less pronounced for small probe times $\Delta^{\ast}$ than in the
case of larger probe times if we relate this route to the $\gamma$ exponent. 
Thus, at larger probe times, survival distributions between postyield and 
\emph{in-strain-hardening} entanglements tend to higher magnitudes more 
``rapidly'' upon deformation. Finally, these trends in the
$\gamma$ regime are also more and more pronounced when
the amount of \emph{loop} chains increases, as it is shown in Fig.~\ref{slopes_surv.fig}(b),
where the "memory" exponent $\gamma$ is plotted against the probe time. 
Consequently, memory effects and long time correlations between sets of mutual 
entanglements are more and more important in the postyield regime, and the
\emph{loop} chains reinforce the memory effects.

In Fig.~\ref{slopes_surv.fig}(a), the increase of $\vert\alpha\vert$ with
$\Delta^{\ast}$ denotes an increase of the loss of the
contribution of preyield mutual entanglements to postyield ones.
Thus, long-time correlations become weaker and weaker between the sets of mutual entanglements
from these two regimes. The values of $\vert\alpha\vert$ for the $100\%$ \emph{loop} triblock are
also higher, especially for $\Delta^{\ast} < t_{p1,p2}$, which points out that this triblock
experiences more chain conformations than others.

Exponents $\alpha$ and $\gamma$ for the polymer glasses are plotted in the insets
of respectively Figs.~\ref{slopes_surv.fig}(a) and (b).
For the isotropic glass, the loss exponent $\alpha$ is higher than for the directed glass.
It is because intermediate conformation states are more accessible for the isotropic glass, 
and thus it can explore more mutual entanglement sets. The slope of
$\vert\alpha\vert$ with respect to $\Delta^{\ast}$ is also lower when compared to triblocks,
because of the pure glassy state that slows down the debonding.

Finally, the memory exponent $\gamma$, which characterizes
the postyield to strain-hardening contributions of sets of mutual entanglements, 
depicts, in Fig.~\ref{slopes_surv.fig}(b), a common trend for both polymer glasses.
With the increase of the probe time $\Delta^{\ast}$, it then appears that 
memory effect and long-time correlations become stronger. This trend
seems to be more pronounced than for triblocks, which is again due to the pure glassy state.

\section{Concluding remarks and outlooks}
\label{sec:Conclude}

During this study, an isotropic, a directed polymer glass and
lamellar directed triblock copolymers were generated. The directed
triblocks were developed in
order to mimic the mechanical behavior of glassy/rubbery/glassy
nanostructurated materials in the strong segregation regime. 
By tuning the amount of \emph{uncrossing}
or \emph{loop} chains, it was possible to simulate different
glassy/rubbery interface definitions. 
Material mechanical properties were investigated using a boundary driven tensile test experiment.

Using this methodology, the impact of \emph{loop} chains on the
mechanical response of triblocks has been approached. Their stress-strain curves depict
large stress drops that come earlier upon deformation, with the increase of the
amount of \emph{loop} chains. A chevron-type transition is also observed, with a strength 
that increases with the same amount. Besides this facts, the global mechanical 
properties seem to not be deeply altered by these amounts, so that, it is hard to predict
a critical amount of \emph{loop} chains above which triblocks integrity is achieved.
To overcome this fact, larger systems with repeated lamellar units should be considered in order
to improve statistics, decrease fluctuations and reach a self-averaging regime.

The sets of mutual entanglements between chains were
monitored during deformation, using a primitive path algorithm. A specific attention has
been paid to the study of their statistics. 

Given the initial undeformed sets of mutual entanglements, the evolution of the ratio 
of created over annihilated events, as well as
the distributions of survival events, has been monitored.
It has been shown that these quantities
follow the two main parts of the stress-strain curve, namely, the preyield and
postyield regimes. Monitoring \emph{on the fly} sets of entanglements during the deformation 
allowed us to also approach the internal metastable cross-link states.

It appears that during the strain-hardening
regime, the debonding of the initial mutual entanglement network occurs.
Debonding of \emph{on the fly} entanglement sets is correlated in time,
depending the initial time window from which depart these sets. If one
considers sets from the preyield regime, their contributions to
the postyield regime decrease with a power-law behavior denoted by a loss
exponent $\alpha$. On the other way, ones that depart from the postyield regime depict
memory effects in the strain-hardening regime. In that case,
a power-law behavior is observed, which is characterized by a memory exponent
$\gamma$. The same trends occur for polymer glasses as well as for triblocks.

For triblocks with a non-null amount of \emph{crossing}
chains, the lower the amount is, the longer
the memory effect of initial undeformed sets of mutual entanglements is. 
On the same way, \emph{loop} chains also enhance long-time correlations and memory effects of internal metastable
cross-link states continuously explored upon deformation.

Finally, it should be noted that the results obtained may depend on the applied strain rate.
Further works should be employed in order to quantify this dependence.

The methodology developed during this study gives rise to further insights.
First, the created and activated sets of mutual entanglements emerge from a complex process, 
which could contribute to new stable intermediate conformations as well as to participate to 
destabilization processes and related unstable states. The inner mechanisms of propagation of these 
ensemble sets should be more precisely studied, and especially how the large stress drops in 
Fig.~\ref{stress_strain.fig} could be interpreted in terms of entanglements and how the amount of
\emph{loop} chains compensates the loss of topological constraints, as observed in 
Fig.\ref{rate_ap-am_init.fig}.
With the methodology developed in this study, one could access these localized or delocalized (long-range) processes. A quasistatic protocol could be suitable in order to remove the thermally activated
processes and distinguish the transmission mechanisms.

Furthermore, this methodology allowed us to deal with the problematic of the way 
external forcing is ``internalized'' by the systems via the cross-link process. 
The rheological response of these systems involves
permanent cross-linked and uncross-linked sets, which can be described by some
surviving probabilities $\zeta(t)$, until $t$ times of deformation occur after cross-linking. 
These states hold internal stresses. Under external
forcing, they are metastable and require far from equilibrium operations in order to be created. 
In the studied systems, which depict rubbery phase or ``interchain'' connectivities,
the generation of internal states as
metastable states is closely related to the notion of plasticity. This
notion is itself related, probably among many others, to some time scale
competitions that take place during this generation process. 
This may lead to an instantaneous momentarily decrease of
time scales, leading to a bias in favor of the future metastable
state. Then, it would be also interesting to extend the notion of entanglement sets 
to the study of internal stress dissipation
mechanisms and local displacement fields.

\end{document}